\documentclass[11pt,preprint]{aastex}
\usepackage{psfig}      

\begin{document}

\title{CONSTRAINING BRANS-DICKE GRAVITY WITH ACCRETING MILLISECOND PULSARS
IN ULTRACOMPACT BINARIES}

\author{Dimitrios Psaltis\altaffilmark{1}}

\affil{Physics and Astronomy Departments, University of
Arizona, 1118 E. 4th St., Tucson, AZ 85721\\
dpsaltis@physics.arizona.edu}

\altaffiltext{1}{also, Sabanci University, Orhanli-Tuzla, 34956 Istanbul,
Turkey}

\begin{abstract}
The five accreting, millisecond X-ray pulsars in ultracompact binaries
that were recently discovered with the {\em Rossi X-ray Timing
Explorer\/} provide excellent candidates for constraining the
deviations from general relativity described by the Brans-Dicke
parameter $\omega_{\rm BD}$. I calculate the expected rate of change
of the orbital periods of these binaries and discuss the prospects of
constraining $\omega_{\rm BD}$ to values that are an order of
magnitude larger than current constraints. Finally, I show how
measurements of the orbital period derivative in ultracompact binaries
can be used to place lower bounds on their orbital inclination.
\end{abstract}

  
\section{INTRODUCTION}

Neutron stars in compact binaries provide some of the best physical
settings for testing the predictions of general relativity. The double
neutron-star systems (the prototypical of which contains the
Hulse-Taylor pulsar) have led to measurements of general relativistic
predictions, such as the periastron precession and the Shapiro delay,
and have given the first indirect evidence for the existence of
gravitational waves (for a recent review see Will 2001).

When compared to solar system tests, however, double neutron star
systems have provided only limited constraints, on alternative
theories of gravity such as the Brans-Dicke theory (Will 2001). This
is mostly due to the fact that the difference in the orbital period
evolution of the binaries between general relativity and Brans-Dicke
gravity is related to the mass difference of the two neutron stars in
each system (Will 2001) and the two members of all double neutron-star
systems have very similar masses (see Thorsett \& Chakrabary 1999).
Moreover, Brans-Dicke gravity can be constructed to be only slightly
different from general relativity, through a single parameter
$\omega_{\rm BD}$, making the constraint on $\omega_{\rm BD}$ rather
weak.

Neutron stars in nature appear in various types of binaries with very
small mass companions, which can in principle be used in placing
stronger constraints on Brans-Dicke gravity. Tens of millisecond radio
pulsars have been discovered in orbits around low mass white dwarfs,
but with orbital separations too large for gravitational radiation to
affect their orbital period evolution (Phinney \& Kulkarni
1994). Neutron stars with low mass companions in close orbits often
appear as bright X-ray sources but their orbital period evolution is
dominated by mass transfer and mass loss from the companion stars in
the form of a magnetic wind (Verbunt 1993). Moreover, most of these
neutron stars show no periodic modulations of their X-ray flux at the
stellar spin frequency, hampering measurements of the orbital periods
and their evolution (Vaughan et al.\ 1994). 

The most compact X-ray binary known to date, 4U~1820$-$30, which
consists of a neutron star in a 11~min orbit around a $\sim 0.067
M_\odot$ companion (Rappaport et al.\ 1987), shows periodic X-ray
eclipses, which were used by Morgan, Remillard, \& Garcia (1988) to
place an upper limit on its orbital period derivative of
$\dot{P}/P<3\times 10^{-7}$~yr$^{-1}$. This limit was subsequently
used by Will \& Zaglauer (1989) to constrain the Brans-Dicke parameter
to $\omega_{\rm BD}\gtrsim 140$ or $600$, depending on the stiffness
of the neutron-star equation of state. However, 4U~1820$-$30 is near
the center of the globular cluster NGC~6624 and hence an apparent
change of its orbital period may be induced by gravitational
acceleration in the potential of the cluster.  This has been given as
a possible reason for the orbital period {\em decrease\/} at a rate of
$\dot{P}/P\simeq -1.1\times 10^{-7}$~yr$^{-1}$ that was later inferred
for this source by Tan et al.\ (1991). Furthermore, changes in the
pattern of X-ray emission and the geometry of the X-ray eclipses may
be responsible for the apparent changes of the orbital period, as
suggested by van der Klis et al.\ (1993), who used a longer baseline
than Tan et al.\ (1991) and measured a less statistically significant
orbital period decrease at a rate of $\dot{P}/P\simeq -5.3\times
10^{-8}$~yr$^{-1}$. Subsequent analysis of archival and more recent
data from {\em RXTE\/} by Chou \& Grindlay (2001) gave a marginal
detection of orbital period decrease at a rate of $\dot{P}/P\simeq
-3.47\times 10^{-8}$~yr$^{-1}$. All these measurements render the
constraint on the Brans-Dicke parameter $\omega_{\rm BD}$ from
4U~1820$-$30 quite uncertain.

Recently, five accreting millisecond pulsars were discovered by {\em
RXTE\/} in very compact binaries with orbital periods between 40~min
and 4.3~hr (SAX~J1808.4$-$3658: Chakrabarty \& Morgan 1998;
XTE~J1751$-$305: Markwardt et al.\ 2002; XTE~J0929$-$3314: Galloway et
al.\ 2002; XTE~J1807$-$294: Markward, Smith, \& Swank 2003;
XTE~J1814$-$338: Markwardt \& Swank 2003). Because the primary stars in
these systems are millisecond pulsars, their orbital periods can be
measured with high accuracy without significant systematic
effects. Their orbital separations are very small and hence angular
momentum losses via magnetic stellar winds are expected to be
minimal. Moreover, all these binaries are in the galactic disk and
thus are not subject to significant gravitational accelerations.
Therefore, the five accreting millisecond pulsars provide prime 
candidates for constraining deviations from general relativity and,
in particular, for constraining the parameter $\omega_{\rm BD}$ of
Brans-Dicke gravity. 

In this article, I follow the analysis of Will \& Zaglauer (1989) to
calculate the constraints on the Brans-Dicke parameter $\omega_{\rm
BD}$ imposed by the measurement of an orbital period derivative for an
accreting millisecond pulsar. I argue that the expected constraints
can be an order of magnitude larger than the constraints from solar
system tests. I finally discuss how a limit on (or measurement of) the
orbital period derivative can be used in placing a lower bound on the
orbital inclinations of the binaries.

\section{ORBITAL EVOLUTION IN BRANS-DICKE GRAVITY}

In this section, I sketch the derivation of the rate of change of the
orbital period of a binary in Brans-Dicke gravity. I follow closely
the analysis of Will \& Zaglauer (1989) but make use of more accurate
relations that describe the evolution of the companion stars.

I consider the evolution of the nearly-circular orbit of a binary
consisting of a neutron star, with mass $m_1$, and a low-mass
companion, with mass $m_2$. The rate of change of the orbital angular
momentum $J=\mu(Gma)$, where $m=m_1+m_2$, $\mu=m_1m_2/m$, and $a$ is
the semi-major axis, is
\begin{eqnarray}
\frac{\dot{J}}{J}&=&\frac{1}{J}\frac{\partial J}{\partial m_1}\dot{m}_1
+\frac{1}{J}\frac{\partial J}{\partial m_2}\dot{m}_2
+\frac{1}{J}\frac{\partial J}{\partial a}\dot{a}\nonumber\\
&=&\left(1-\frac{\beta}{q}-\frac{1}{2}\frac{1-\beta}{1+q}\right)
   \frac{\dot{m}_2}{m_2}+\frac{1}{2}\left(\frac{\dot{a}}{a}\right)\;,
\label{jdot}
\end{eqnarray}
where $q\equiv m_1/m_2$ is the mass ratio in the binary and
$\beta=-\dot{m}_1 /\dot{m}_2$ is a parameter that describes the rate
of mass loss from the system, if the mass-transfer is non-conservative.

For systems such as the newly discovered ultracompact binaries,
orbital angular momentum may be lost because of gravitational-wave
radiation or mass loss from the systems. Therefore, the left-hand-side
of equation~(\ref{jdot}) is equal to
\begin{equation}
\frac{\dot{J}}{J}=\frac{\dot{J}_{\rm rad}}{J}+j_{\rm w}(1-\beta)
\frac{1+q}{q}\frac{\dot{m_2}}{m_2}\;,
\label{jdot2}
\end{equation}
where $\dot{J}_{\rm rad}$ is the rate of loss of angular momentum
caused by gravitational-wave radiation and $j_{\rm w}$ is the specific
angular momentum carried away by the stellar wind, in units of $2\pi a^2/P$,
where $P$ is the orbital period.

I am interested in comparing the predictions of different gravity
theories to the observed rate of change of the orbital period, which I
calculate using (e.g., Will \& Zaglauer 1989)
\begin{equation}
\frac{P}{2\pi}=\frac{m}{m_1^3 m_2^3}J^3 G^{-2}
\end{equation}
so that
\begin{equation}
\frac{\dot{P}}{P}=\left(\frac{1-\beta}{1+q}-
\frac{3}{2}\frac{1-\beta}{1+q}\right)\frac{\dot{m}_2}{m_2}
+\frac{3}{2}\frac{\dot{a}}{a}\;.
\label{Pdot1}
\end{equation}

In order to evaluate the rate of change of the semi-major axis, I will
assume that the companion star always fills its Roche lobe, i.e., that
its radius $R_2$ is equal to (Eggleton 1983)
\begin{equation}
R_2=\frac{0.49 q^{-2/3}}{0.6q^{-2/3}+\ln(1+q^{-1/3})}a\;.
\end{equation}
Using the fact that 
\begin{equation}
\frac{\dot{q}}{q}=-\left(\frac{\beta+q}{q}\right)\frac{\dot{m}_2}{m_2}\;,
\end{equation}
I can evaluate the rate of change of the companion's radius to be
\begin{equation}
\frac{\dot{R}_2}{R_2}=\frac{\dot{a}}{a}+\frac{2}{3}\left(\frac{\beta+q}{q}
\right)
\left[1-\frac{0.6+0.49q^{1/3}(1+q^{-1/3})^{-1}}{0.6+q^{2/3}\ln(1+q^{-1/3})}
\right]\frac{\dot{m}_2}{m_2}\;.
\end{equation}
I can then turn the last equation into
\begin{equation}
\frac{\dot{a}}{a}=\left\{\xi_{\rm ad}-
\frac{2}{3}\left(\frac{\beta+q}{q}
\right)
\left[1-\frac{0.6+0.49q^{1/3}(1+q^{-1/3})^{-1}}{0.6+q^{2/3}\ln(1+q^{-1/3})}
\right]\right\}\frac{\dot{m}_2}{m_2}\;,
\label{adot}
\end{equation}
where I have introduced the adiabatic index $\xi_{\rm ad}\equiv d\ln R_2/
d\ln m_2$ for the companion star.

I now combine equations~(\ref{jdot}), (\ref{jdot2}), (\ref{Pdot1}), and
(\ref{adot}) to obtain
\begin{equation}
\frac{\dot{P}}{P}=3\left(\frac{n}{D}\right)\frac{\dot{J}_{\rm rad}}{J}\;,
\label{Pdot}
\end{equation}
where 
\begin{equation}
n\equiv\frac{1}{2}\left(\xi_{\rm ad}-\frac{1}{3}\frac{1-\beta}{1+q}
-\frac{2}{3}\frac{\beta+q}{q}{\cal A}\right)\;,
\end{equation}
\begin{equation}
D\equiv 1+\frac{1}{2}\xi_{\rm ad}-\frac{1}{2}\left(\frac{1-\beta}{1+q}\right)-
\frac{1}{q}\left[\beta+j_w(1-\beta)(1+q)\right]-\frac{1}{3}\frac{\beta+q}{q}
{\cal A}\;,
\end{equation}
and 
\begin{equation}
{\cal A}\equiv \left[1-\frac{0.6+0.49q^{1/3}(1+q^{-1/3})^{-1}}{0.6+q^{2/3}\ln(1+q^{-1/3})}
\right]\;.
\end{equation}

Equation~(\ref{Pdot}) is more general than the one derived by Will \&
Zaglauer (1989), as it is valid for a wider range of mass ratios of
the binary. It provides the predicted rate of change of the orbital
period of a binary, given a rate of angular momentum lost by the
emission of gravitational waves and by mass loss from the binary. For
a Brans-Dicke gravity, this rate is equal to (Will \& Zaglauer 1989)
\begin{equation}
\frac{\dot{J}_{\rm rad}}{J}=-\frac{1}{3}\left[\frac{96}{5}\frac{\mu m^2}{a^4}
\left(\frac{k_1}{12}\right)+\frac{2\mu m}{a^3}{\cal G}\xi s_1^2\right]\;,
\end{equation}
where
\begin{equation}
{\cal G}\equiv 1-\xi s_1\;,
\end{equation}
\begin{equation}
k_1={\cal G}^2\left[12\left(1-\frac{1}{2}\xi\right)+\xi
 \left(1-2\frac{m_1 s_1}{m_1+m_2}\right)\right]\;,
\end{equation}
\begin{equation}
s\equiv -\left.\frac{\partial \ln m_1}{\partial G}\right\vert_0
\end{equation}
is the so-called neutron-star sensitivity, and
\begin{equation}
\xi\equiv\frac{1}{2+\omega_{\rm BD}}
\label{xi}
\end{equation}
is a parameter that describes the deviation of Brans-Dicke gravity
from general relativity. Note that general relativity corresponds to
$\omega_{\rm BD}\rightarrow \infty$ and hence to $\xi=0$. In writing
the above equations, I assumed that the companion to the neutron star
is a non-relativistic star and hence the sensitivity of the latter is
negligible.

Given a rate of change of the orbital period of the binary
($\dot{P}/P$), the orbital parameters of the binary (i.e., $m_1$,
$m_2$, $P$, and $a$), the properties of the neutron star (i.e., the
value of $s$) and of the companion star (i.e., the value of $\xi_{\rm ad}$),
and the properties of mass loss in the binary (i.e., the values of
$\beta$ and $j_{\rm w}$), equations~(\ref{Pdot})-(\ref{xi}) can be
used to place a constraint on the Brans-Dicke parameter $\omega_{\rm
BD}$.

\section{APPLICATION TO ULTRACOMPACT BINARIES}

In section \S2, I sketched (following Will \& Zaglauer 1989) the
derivation of the rate of change of the orbital period of an
ultracompact binary, when orbital angular momentum is lost due to
emission of gravitational radiation, in Brans-Dicke gravity. In this
section, I will discuss the observed properties of the recently
discovered ultracompact binaries and their prospect for constraining
deviations of this theory from general relativity.

\noindent {\em Properties of the binaries.---\/}Five accreting
millisecond pulsars have been discovered to date, in ultracompact
binaries with orbital periods between 40~minutes and 4.3 hours.  Their
orbital periods $P_{\rm orb}$, projected semi-major axes $a$,
eccentricities $e$, and mass functions $f$ are summarized in Table~1.

\begin{table}[t]
\centerline{
\footnotesize
\begin{tabular}{lcccl}
\multicolumn{5}{c}{Table 1: Observed Properties of Ultracompact Binaries}\\
\hline
Source & $P_{\rm orb}$ (min) & $a$ (lt-ms) & $f$~$(M_\odot)$ & Reference \\
\hline
SAX~J1808.4$-$3658 & 120.9 & 62.809
    & $3.78\times 10^{-5}$ & Chakrabarty \& Morgan 1998\\
XTE~J0929$-$3314 & 43.6 & 6.290 
    & $2.7\times 10^{-7}$ & Galloway et al.\ 2002\\
XTE~J1751$-$305 & 42.4 & 10.1134 
    & $1.278\times10^{-6}$ & Markwardt et al.\ 2002\\
XTE~J1807$-$294 & 40.1 & 4.80 & $1.54\times 10^{-7}$ 
    &  Markwardt et al.\ 2003; Markwardt priv.\ comm.\\
XTE~J1814$-$338 & 256.5 & 390.3 & $2.016\times 10^{-3}$ 
    & Markwardt \& Swank 2003; Markwardt priv.\ comm.\\
\hline
\end{tabular}}
\end{table}

The constraint on the parameter $\omega_{\rm BD}$ depends on the
measured orbital period and semi-major axis of each orbit, as well as
on the mass of the neutron star and of the companion star. A
combination of observational and theoretical arguments strongly
constrain the mass of the former to lie in the narrow range $\simeq
1.3-2.2 M_{\odot}$ (see, e.g., Lattimer \& Prakash 1999). The mass of
the companion star is then calculated as a function of the unknown
inclination $i$ of the binary using the mass function, i.e.,
\begin{equation}
f=\frac{(m_2 \sin i)^3}{(m_1+m_2)^2}\;.
\label{f}
\end{equation}
For reasons related to the stability of mass transfer in X-ray binary
systems, I will only consider cases in which the companion star is
less massive than the neutron star.

For the parameters of the ultracompact binaries discussed here, the
minimum absolute value of the rate of change of the orbital period
increases with increasing mass of the neutron star or the companion
star.  This is caused by the fact that both the semi-major axis of a
binary of given period and the rate of emission of gravitational waves
increase with increasing total mass of the system. As
equation~(\ref{f}) shows, the limiting absolute value for the rate of
change of the orbital period corresponds to the maximum inclination (i.e,
$\sin i=1$) and the minimum accepted value for the neutron-star mass, which
we take to be 1.3~$M_\odot$.

\begin{figure}[t]
 \centerline{
\psfig{file=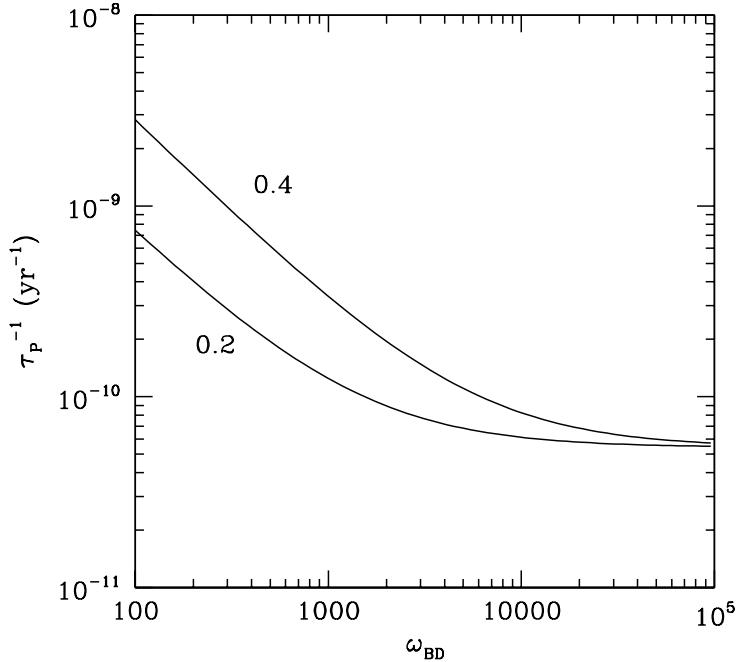,angle=0,width=10truecm}}
\figcaption[]{\footnotesize The rate of orbital period
evolution ($\tau_{\rm P}^{-1}\equiv \dot{P}/P$) as a function of the
Brans-Dicke parameter $\omega_{\rm BD}$, for different values of the
neutron-star sensitivity $s$. For this plot, I used the orbital
parameters of XTE~J1808.4$-$3658 and assumed a neutron star mass of
$1.3 M_\odot$, no mass loss, an edge-on orbit, and $\xi_{\rm ad}=0$ for
the companion star.
\label{fig:NSsens}}
\end{figure}

\noindent {\em Properties of the neutron stars.---\/}The
separations of the ultracompact binaries under study are much larger
than the radii of the neutron stars. Therefore, in general relativity,
the specifics of the neutron-star structure, and hence its equation of
state, do not enter in the calculation of the angular momentum lost
due to gravitational radiation. However, this is not true in
Brans-Dicke gravity, in which the self-gravitational binding energies
of the neutron stars (i.e., their sensitivities $s$) affect the result
(Will \& Zaglauer 1989).

A number of values for the sensitivity have been calculated
by Will \& Zaglauer (1989) and by Zaglauer (1990) using the relation
\begin{equation}
s\equiv=-\left(\frac{\partial \ln m_1}{\partial \ln G}\right)_N=\frac{3}{2}
\left[1-\left(\frac{\partial \ln m_1}{\partial \ln N}\right)_G\right]\;,
\end{equation}
where $N$ is the baryonic number in the star. According to these
calculations, the sensitivity increases with increasing mass and with
the stiffness of the equation of state, as both make the stars more
compact and hence increase their self-gravitational binding energies.

The minimum absolute value of the rate of change of the orbital period
also increases with increasing neutron star sensitivity, as shown for
a typical case in Figure~\ref{fig:NSsens}. The most stringent limit on
the Brans-Dicke parameter $\omega_{\rm BD}$ will, therefore, be obtained
when using the smallest sensitivity for the lightest neutron star
considered. For the rest of the paper, I will assume this to be equal
to 0.2.

\begin{figure}[t]
 \centerline{
\psfig{file=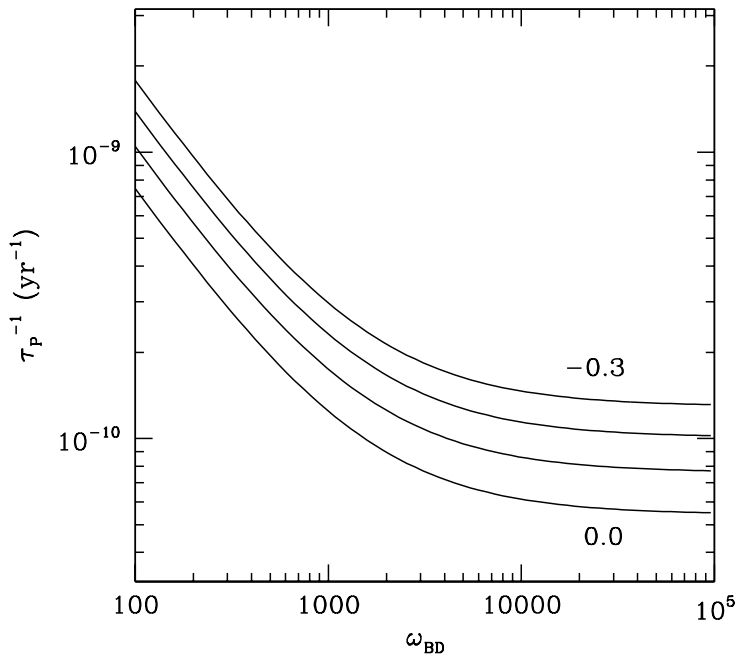,angle=0,width=10truecm}}
\figcaption[]{\footnotesize The rate of orbital period
evolution ($\tau_{\rm P}^{-1}\equiv \dot{P}/P$) as a function of the
Brans-Dicke parameter $\omega_{\rm BD}$, for different values of the
adiabatic index $\xi_{\rm ad}$ of the companion star. For this plot, I
assumed that the neutron star sensitivity is equal to 0.2 and used the
same parameters as in Fig.~\ref{fig:NSsens}.
\label{fig:ksiad}}
\end{figure}

{\em Properties of the companion stars.---\/}The response of the
companion star to mass loss is described by the adiabatic index
$\xi_{\rm ad}$ and depends on the nature of the star as well as on a
number of factors such as the presence of external irradiation and the
rate of mass loss. If all external effects are neglected and the star
is considered to be a polytrope, the parameter $\xi_{\rm ad}$ takes the
canonical value of $\xi_{\rm ad}=-1/3$.

The companions to the neutron stars in the ultracompact binaries under
consideration are believed to be white dwarfs (see Deloye \& Bildsten
2003 and references therein), given their inferred very low masses and
small sizes (both are of order a few hundredths of the solar mass and
radius, respectively). The response of such stars to adiabatic mass
loss was recently calculated in detail by Deloye \& Bildsten (2003), who
found that the corresponding parameter $\xi_{\rm ad}$ decreases with
increasing mass of the white dwarf and with decreasing atomic weight
of the main compositional element, but depends rather weakly on its
temperature. For the inferred values of the masses of the white
dwarfs in the ultracompact binaries, they showed that $-0.3\lesssim
\xi_{\rm ad}\lesssim 0$.

The dependence of the rate of orbital period evolution on the
parameter $\xi_{\rm ad}$ is shown in Figure~\ref{fig:ksiad}. Clearly, for a
weaker response of the companion star (i.e., for smaller values of
$\vert\xi_{\rm ad}\vert$), the rate of orbital period change is also
smaller. This is expected, given our assumption that the companion
star always fills its Roche lobe for mass transfer to occur and,
therefore, a small change in the radius of the star caused by mass
loss will be accompanied by a small change in the orbital
period. Because the weakest companion-star response corresponds to the
lowest mass stars (see Deloye \& Bildsten 2003) and the latter also
produce the lowest rate of orbital period change, using the lowest
allowed value of $\vert\xi_{\rm ad}\vert$ for the lowest companion mass in each
binary will provide the most stringent limit.

\begin{figure}[t]
 \centerline{
\psfig{file=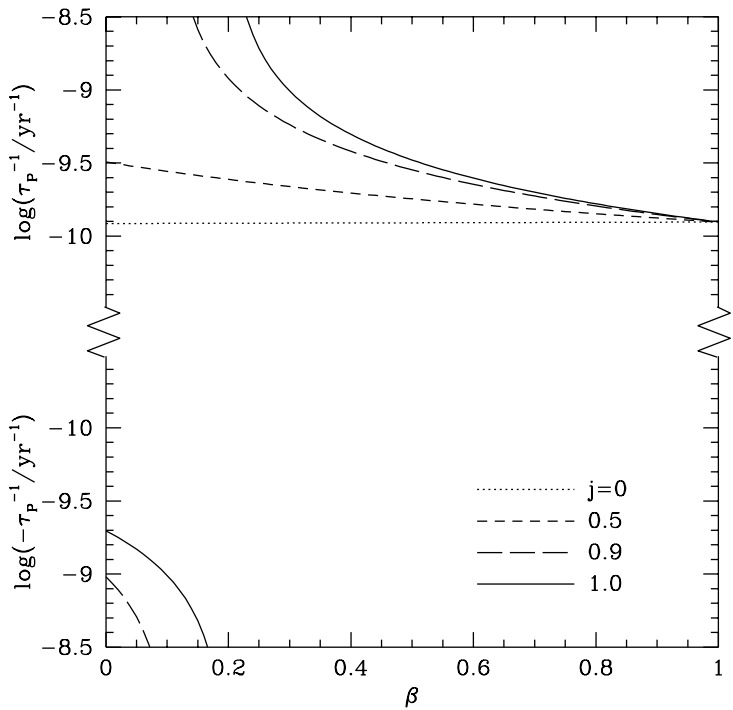,angle=0,width=10truecm}}
\figcaption[]{\footnotesize The rate of  
orbital period evolution ($\tau_{\rm P}^{-1}\equiv \dot{P}/P$) as a
function of the mass-loss parameter $\beta$, for different values of
the angular momentum of the wind, $j_{\rm w}$. The binary parameters
are the same as in Figure~1, the neutron-star sensitivity is set to
$s=0.2$, the adiabatic index for the companion star is set to
$\xi_{\rm ad}=0$ and the Brans-Dicke parameter is set to $\omega_{\rm
BD}=1000$. The lower half of the plot corresponds to an orbital period
that decreases with time ($\dot{P}/P<0$), whereas the upper half
corresponds to an orbital period that increases with time
($\dot{P}/P>0$).
\label{fig:mloss}}
\end{figure}

{\em Properties of the mass loss.---}The last and hardest to constrain
uncertainty in the calculation of the predicted rate of orbital period
change arises from the properties of the mass loss from the binary,
i.e., the parameters $\beta$ and $j_{\rm w}$. Even though the
companions to the neutron stars in the ultracompact binaries are white
dwarfs and lie deep in the gravitational potential wells of the neutron
stars, mass from their surfaces may be lost from the binary systems
because of the ablation caused by the intense X-ray irradiation.

The dependence of the predicted rate of change of the orbital period
on the amount ($\beta$) and strength ($j_w$) of mass loss is shown in
Figure~\ref{fig:mloss}. Clearly, when there is very little mass loss,
i.e., when $\beta\rightarrow 1$, the orbital period increases with
time and the predicted rate depends very weakly on $j_{\rm w}$. The
situation is very different, however, when most of the mass is lost
from the system, i.e., when $\beta\rightarrow 0$.

In order to understand this dependence, I rewrite
equation~(\ref{Pdot}) taking the limit $q\gg 1$, which is appropriate
for the ultracompact binaries under study. In this case, ${\cal
A}\rightarrow 1/2$ and
\begin{equation}
\frac{\dot{P}}{P}\simeq 3 \left[\frac{\xi_{\rm ad}-\frac{1}{3}}
   {\frac{5}{3}+\xi_{\rm ad}-
    j_{\rm w}(1-\beta)}\right]\frac{\dot{J}_{\rm rad}}{J}\;.
\label{Pdotaprx}
\end{equation}
In the absence of mass loss, equation~(\ref{Pdotaprx}) shows that,
even though angular momentum is lost from the companion star due to
the emission of gravitational waves, i.e., $\dot{J}_{\rm rad}<0$, the
period of the binary increases, as long as $\xi_{\rm ad}<1/3$. This is a direct
consequence of the assumption that the companion star always fills its
Roche lobe for mass transfer to occur and of the fact that the radii
of these low-mass white dwarfs increase with decreasing mass. However,
if a significant amount of angular momentum is removed by the wind in
addition to the gravitational radiation, i.e., if
\begin{equation}
j_{\rm w}(1-\beta)>\frac{5}{3}+\xi_{\rm ad}\;,
\end{equation}
then the orbit of the binary shrinks and the orbital period decreases.

It is important to note, however, that, whether there is mass loss or
not, there is always a lower limit on the absolute value of the rate
of change of the orbital period, since
\begin{equation}
\left\vert\frac{\dot{P}}{P}\right\vert\simeq 
\frac{1-3\xi_{\rm ad}}{\vert\frac{5}{3}+\xi_{\rm ad}+j_{\rm w} \beta-j_{\rm w}\vert}
\left\vert\frac{\dot{J}_{\rm rad}}{J}\right\vert \ge
\frac{1-3\xi_{\rm ad}}{\frac{5}{3}+\xi_{\rm ad}+j_{\rm w} \beta+j_{\rm w}}
\left\vert\frac{\dot{J}_{\rm rad}}{J}\right\vert 
\label{Pdotmin}
\end{equation}
Equation~(\ref{Pdotaprx}) shows that the minimum positive rate of
change of the orbital period corresponds to the limit
$\beta\rightarrow 1$, whereas the maximum negative rate of change of
the orbital period corresponds to the limit $\beta \rightarrow 0$ and
$j_{\rm w}\rightarrow 1$ (see also Fig.~\ref{fig:mloss}). Moreover,
the absolute values of these two limits are comparable in size. This
fact allows for stringent constraints to be placed on the Brans-Dicke
parameter $\omega_{\rm BD}$ by the measurement of an orbital period
derivative, independent of whether the latter has a positive or
negative value.

\section{RESULTS AND DISCUSSION}

In \S3, I studied the rate of orbital period evolution in an
ultracompact binary and its dependence on the various model
parameters.  When the binary period is increasing with time, I found
that the rate of evolution is minimized for the lowest neutron-star
mass (taken here to be $\ge 1.3M_\odot$), the largest inclination
($\sin i=1$), the largest adiabatic index for the companion star (take
here to be $\xi_{\rm ad}\le0$), and the case of no mass-loss
($\beta=1$). On the other hand, when the binary period is decreasing
with time, the absolute value of this rate is minimized for the same
stellar parameters as before but at the limit of complete mass loss
($\beta=0$) and with the wind carrying the orbital angular momentum
($j_{\rm w}=1$).

\begin{figure}[t]
 \centerline{
\psfig{file=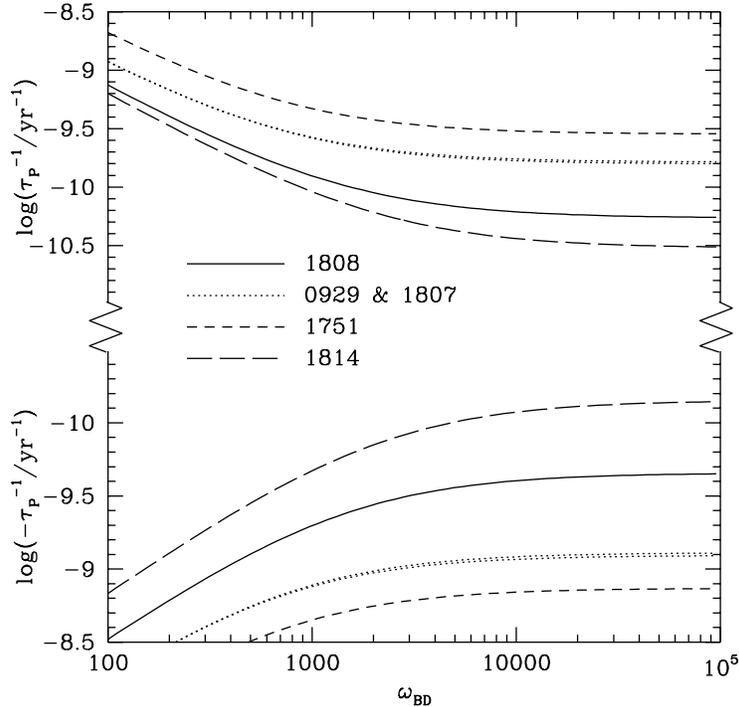,angle=0,width=10truecm}}
\figcaption[]{\footnotesize The limiting rate for the evolution of the 
orbital period ($\tau_{\rm P}^{-1}\equiv \dot{P}/P$) as a function of the
Brans-Dicke parameter $\omega_{\rm BD}$, for the orbital parameters of
the five known millisecond accreting pulsars. The lower half of the
plot corresponds to an orbital period that decreases with time
($\dot{P}/P<0$), whereas the upper half corresponds to an orbital
period that increases with time ($\dot{P}/P>0$). \label{fig:constr}}
\end{figure}

Figure~\ref{fig:constr} shows the limiting rate of orbital period
evolution as a function of the Brans-Dicke parameter $\omega_{\rm
BD}$, for the five known accreting millisecond pulsars. For each
source, the region between the two corresponding curves is
excluded. As a result, a measurement of either a positive or a
negative rate of orbital period evolution can be used in placing a
lower bound on the Brans-Dicke parameter.

The limiting curves become insensitive to the deviation of Brans-Dicke
gravity from general relativity for $\omega_{\rm BD}\gtrsim 10^4$.
This number represents the tightest constraint that can be achieved
with this method and is approximately an order of magnitude larger
than the constraints imposed by solar-system and double-neutron-star
tests (Will 2001). In deriving these constraining curves, I assumed no
prior knowledge of any of the binary parameters other than the ones
that can be inferred from X-ray timing. These constraints can be
improved by measuring the masses of the neutron star and the companion
star, by identifying the nature of the companions (and hence
constraining their adiabatic indices $\xi_{\rm ad}$), and by placing
constraints on the mass loss from the systems. The tightest limits can
be achieved for an eclipsing ultracompact binary, with a relatively
long orbital period, and limited mass loss.

\begin{figure}[t]
 \centerline{ 
\psfig{file=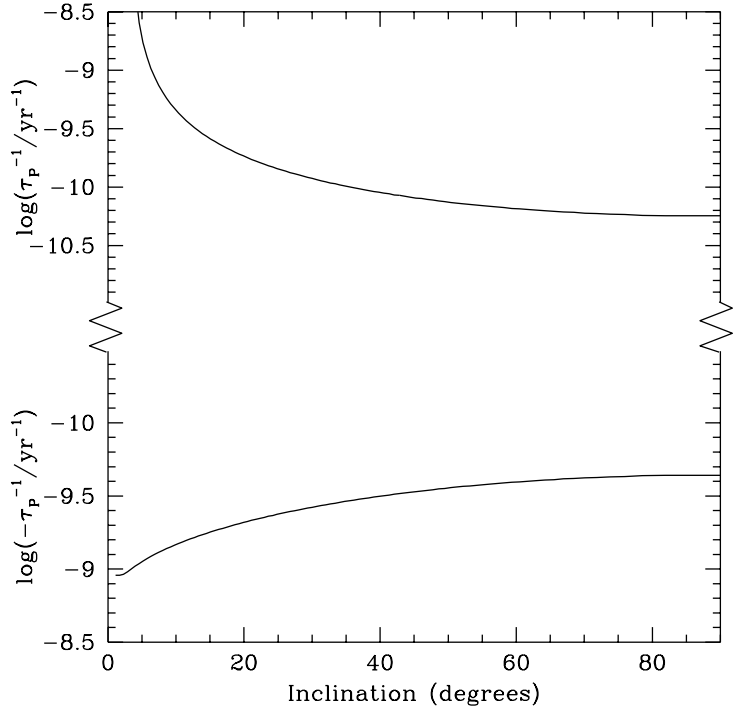,angle=0,width=10truecm}}
\figcaption[]{\footnotesize The limiting rate for the evolution of the 
orbital period ($\tau_{\rm P}^{-1}\equiv \dot{P}/P$) of
SAX~J1808.4$-$3658, in general relativity, as a function of the
orbital inclination.  The lower half of the plot corresponds to an
orbital period that decreases with time ($\dot{P}/P<0$), whereas the
upper half corresponds to an orbital period that increases with time
($\dot{P}/P>0$). \label{fig:incl}}
\end{figure}

Even when the binary parameters are not favorable, and hence the
resulting constraints on the Brans-Dicke parameter $\omega_{\rm BD}$
are not stringent, the analysis presented here can be used in
constraining the orbital inclination of the ultracompact binaries and
the properties of mass loss. As an example, Figure~\ref{fig:incl}
shows the limiting rate of evolution of the orbital period of the
source SAX~J1808.4$-$3658, in general relativity, as a function of the
orbital inclination. The region between the two curves is not allowed
for any neutron-star mass in the range $1.3-2.2 M_{\odot}$ and for any
mass-loss mechanism from the binary. As a result, a measurement of the
rate of orbital period evolution can also be used in placing a lower
bound on the inclination of the binary system.

The measurements required to place a stringent constraint on the
Brans-Dicke parameter $\omega_{\rm BD}$ can be achieved using the
Rossi X-ray Timing Explorer by measuring the orbital periods of the
binaries over a long period of time, i.e., between successive
outbursts. Indeed, the source SAX~J1808.4$-$3658 has shown four
outbursts within six years and the accuracy of the measurement of the
orbital period in the 1998 outburst alone is comparable to the value
needed to achieve a useful constraint (see Chakrabarty \& Morgan
1998). However, the properties of mass loss from the binary system may
not remain constant over a long period of time. If the mass loss is
driven by irradiation of the companion star, then during an outburst
the mass loss will be significant and the orbital period may be
decreasing with time, whereas in between outbursts the mass loss may be
negligible and the orbital period will be increasing with time. The
net result will be an artificially reduced rate of orbital period
evolution and thus an artificially stringent constraint on Brans-Dicke
gravity. The effects of a variable rate of mass loss can be minimized
if an orbital period derivative can be measured in a single, long
outburst. Given the short durations of the outbursts so far observed
from the five known sources, such a measurement is unlikely with the
capabilities of RXTE. It can be, however, one of the key scientific
goals of the next X-ray timing mission (see, e.g., Markwardt 2004).

\acknowledgements
It is a pleasure to thank Duncan Galloway for help in compiling the
data in the table, Deepto Chakrabarty for useful discussions, and
Feryal \"Ozel for a critical reading of the manuscript.  I am also
grateful for their hospitality to the Astrophysics Group at Sabanci
University, in Istanbul, Turkey, where this work was completed.

{\em Note added.---\/}While this work was in its final stages, a new
accreting millisecond pulsar, IGR~J00291$+$5934, was discovered in an
ultracompact binary (Markwardt et al.\ 2004, ATEL \#353, \#360). This
new source has very similar orbital parameters with the five previously
known accreting millisecond pulsars and can also be used in placing 
constraints on Brans-Dicke gravity.

\end{document}